\documentclass[preprint]{aastex63}

\usepackage{enumitem,amssymb,multirow}
\newlist{todolist}{itemize}{2}
\setlist[todolist]{label=$\square$}
\usepackage{pifont}

\usepackage{soul}
\usepackage{colortbl}

\graphicspath{{./}{figures/}}

\received{April 30th, 2020}
\revised{July 29th, 2020}
\accepted{July 31st, 2020}
\submitjournal{PSJ}

\shorttitle{A Search for Stable Venus Co-Orbital Asteroids}
\shortauthors{Pokorn\'{y} and Kuchner}

\begin{document}

\title{A Deep Search for Stable Venus Co-Orbital Asteroids: Limits on The Population}

\correspondingauthor{Petr Pokorn\'{y}}
\email{petr.pokorny@nasa.gov}

\author[0000-0002-5667-9337]{Petr Pokorn\'{y}}
\affiliation{Department of Physics, The Catholic University of America, Washington, DC 20064, USA}
\affiliation{Heliophysics Science Division, NASA Goddard Space Flight Center, Greenbelt, MD 20771}
\affiliation{Astrophysics Science Division, NASA Goddard Space Flight Center, Greenbelt, MD 20771}
\author[0000-0002-2387-5489]{Marc J. Kuchner}
\affiliation{Astrophysics Science Division, NASA Goddard Space Flight Center, Greenbelt, MD 20771}
\author[0000-0003-3145-8682]{Scott S. Sheppard}
\affiliation{Earth and Planets Laboratory, Carnegie Institution for Science, Washington, DC, 20015}

\begin{abstract}
A stable population of objects co-orbiting with Venus was recently hypothesized in order to explain the existence of Venus's co-orbital dust ring. 
We conducted a 5 day twilight survey for these objects with the Cerro-Tololo Inter-American Observatory (CTIO) 4 meter telescope covering about 35 unique square degrees to 21 mag in the $r$-band.
Our survey provides the most stringent limit so far on the number of Venus co-orbital asteroids; it was capable of detecting 5\% of the entire population of those asteroids brighter than 21 magnitude.
\deleted{From this, we estimate an upper limit on the number of co-orbital asteroids brighter than 21 magnitude (approximately {400--900} m in diameter depending on the asteroid albedo) to be {$18^{+30}_{-14}$} asteroids. The mass of the observed co-orbital dust ring was estimated from models to be equivalent to an asteroid with a 2 km diameter ground to dust. Therefore this upper limit suggests that either Venus's co-orbitals are non-reflective at the observed phase angles, have very low albedo ($<1\%$), or that an alternative explanation for the observed dust ring might be needed.}
\added{We estimate an upper limit on the number of co-orbital asteroids brighter than 21 magnitude (approximately 400--900 m in diameter depending on the asteroid albedo) to be {$N=18^{+30}_{-14}$}. Previous studies estimated the mass of the observed dust ring co-orbiting with Venus to be equivalent to an asteroid with a 2 km diameter ground to dust. Our survey estimates $<6$ asteroids larger than 2 km. This implies the following possibilities: that Venus co-orbitals are non-reflective at the observed phase angles, have a very low albedo ($<1\%$), or that the Venus co-orbital dust ring has a source other than asteroids co-orbiting Venus.}
We discuss this result, and as an aid to future searches, we provide predictions for the spatial, visual magnitude, and number density distributions of stable Venus co-orbitals based on the dynamics of the region and magnitude estimates for various asteroid types.

%
\end{abstract}

\section{Introduction}
There are currently 5 known Venus co-orbital asteroids, defined as objects in 1:1 mean motion resonance (MMR) with Venus: 2001 CK$_{32}$ \citep{Brasser_etal_2004}, 2002 VE$_{68}$ \citep{Mikkola_etal_2004}, 2012 XE$_{133}$ \citep{delaFuenteMarcos_delaFuenteMarcos_2013}, 2013 ND$_{15}$ \citep{delaFuenteMarcos_delaFuenteMarcos_2014}, and 2015 WZ$_{12}$ \citep{delaFuenteMarcos_delaFuenteMarcos_2017}. Each of these asteroids has a highly eccentric orbit crossing the orbit of Earth. This known high eccentricity co-orbital population is dynamically unstable on {million-to-billion} year timescales  \citep{Morais_Morbidelli_2006,Christou_2000}.  

Observations also suggest that Venus shepherds a second population of small bodies in lower eccentricity orbits.  Data from two space missions, \textit{HELIOS} \citep{Leinert_Moster_2007} and \textit{STEREO} \citep{Jones_etal_2013}, indicate the presence of a narrow ring of dust in Venus's orbit.  
A recent study investigating the origin of this co-orbital dust ring suggests that asteroids on lower eccentricity orbits ($e < 0.3$) in the 1:1 MMR with Venus are the only possible explanation for the dust ring \citep{Pokorny_Kuchner_2019}. These asteroids on lower eccentricity orbits are understood to be dynamically stable \citep{Cuk_etal_2012, Pokorny_Kuchner_2019}.  \citet{Pokorny_Kuchner_2019} showed that approximately 8\% of Venus co-orbitals remain in a stable 1:1 mean-motion resonance with Venus for the age of the solar system. 

However, this hypothesized population of stable Venus co-orbitals on low-eccentricity orbits has yet to be detected. Observing asteroids, like these low-eccentricity Venus co-orbitals, within 1 au from the Sun is a challenge for both ground-based and space-borne telescopes due to increased sky brightness and higher airmasses {during twilight} \citep[ground-based telescopes;][]{Wiegert_etal_2000} and generally large solar avoidance angles \citep[{e.g., }Hubble Space Telescope, $\lambda>50^\circ$;][]{Mobasher_2002}. \citet{SHEPPARD_TRUJILLO_2009} probed Venus's Hill sphere for the existence of any satellites, without any success, and concluded that Venus probably lacks any natural satellites larger than 500 m. \citet{SHEPPARD_TRUJILLO_2009} also conducts a review of observations of the inner solar system existing at the time, that are able to detect large asteroids and planetary moons, and find no Venus co-orbitals. Recently, \citet{Ye_etal_2020} summarized a search for the innermost solar-system objects using the Zwicky Transient Facility (ZTF) camera at Palomar Observatory.  The survey achieved a limiting magnitude in the r-band of 19.5 at solar elongations down to $35^\circ$, and found no co-orbitals of Venus. \citet{Ye_etal_2020} concluded that if they exist they must have diameters smaller than their limiting magnitude threshold ($D<0.5-1.5$ km; assuming albedos between 0.45 and 0.04) and would require studies with deeper reach in the visual magnitude.

We describe here a new search for stable Venus co-orbital asteroids. We compare the results of our search to a new model for the distribution of these small bodies, and we derive new upper limits on their number.  We also compare the \citet{SHEPPARD_TRUJILLO_2009} survey to our model, and provide recommendations on the best places to continue the search for these objects.

\section{Observing campaign with CTIO 4 meter telescope using Dark Energy Camera}

On the nights of UT September 23, 25, 26, 27, and 28 of 2019 we conducted a survey for Venus co-orbitals using the Dark Energy Camera (DECam) mounted on the 4-meter aperture telescope at Cerro Tololo
Inter-American Observatory, Chile (CTIO). The summary of the observed fields and times of observations is shown in Table \ref{TABLE:CTIO}. DECam has an effective area of CCD focal plane of 3.0 square degrees, which
translates to a circle with a 0.977 degree radius\footnote{\url{{http://www.ctio.noao.edu/noao/node/1033}}}, giving an effective pixel scale of about 0.268 arcseconds \citep{Flaugher_etal_2015}. The survey was conducted in the $r$-band with a limiting magnitude of 21 magnitudes using 20 second exposures with seeing around 0.9 to 1.0 arcseconds.

Four or five fields were imaged each night near the orbit of Venus at the beginning of astronomical twilight.  This is when the sky is relatively dark and yet Venus' orbital plane is still above 2.3 airmasses on the sky at this time of year. Each field was imaged twice on a given night with about a 4 minute interval between images.  This allowed us to see any solar system bodies that had relatively fast apparent motions, as would be expected for any objects near Venus' orbit.  We searched all observed fields for objects moving faster than 90 arcseconds per hour or over 20 pixels of motion (the mean combined drift in RA and Dec from our Venus co-orbital motion was around 130 arcseconds per hour) using a modified algorithm based on \citet{Trujillo_Jewitt_1998} and used for the outer Solar System surveys \citep{Sheppard_etal_2019}. Objects moving slower than 90 arcseconds per hour are expected to be further away than Venus' orbit and mostly be background main belt asteroids, so they were ignored.

\begin{figure}
\begin{center}
\includegraphics[width=0.5\textwidth]{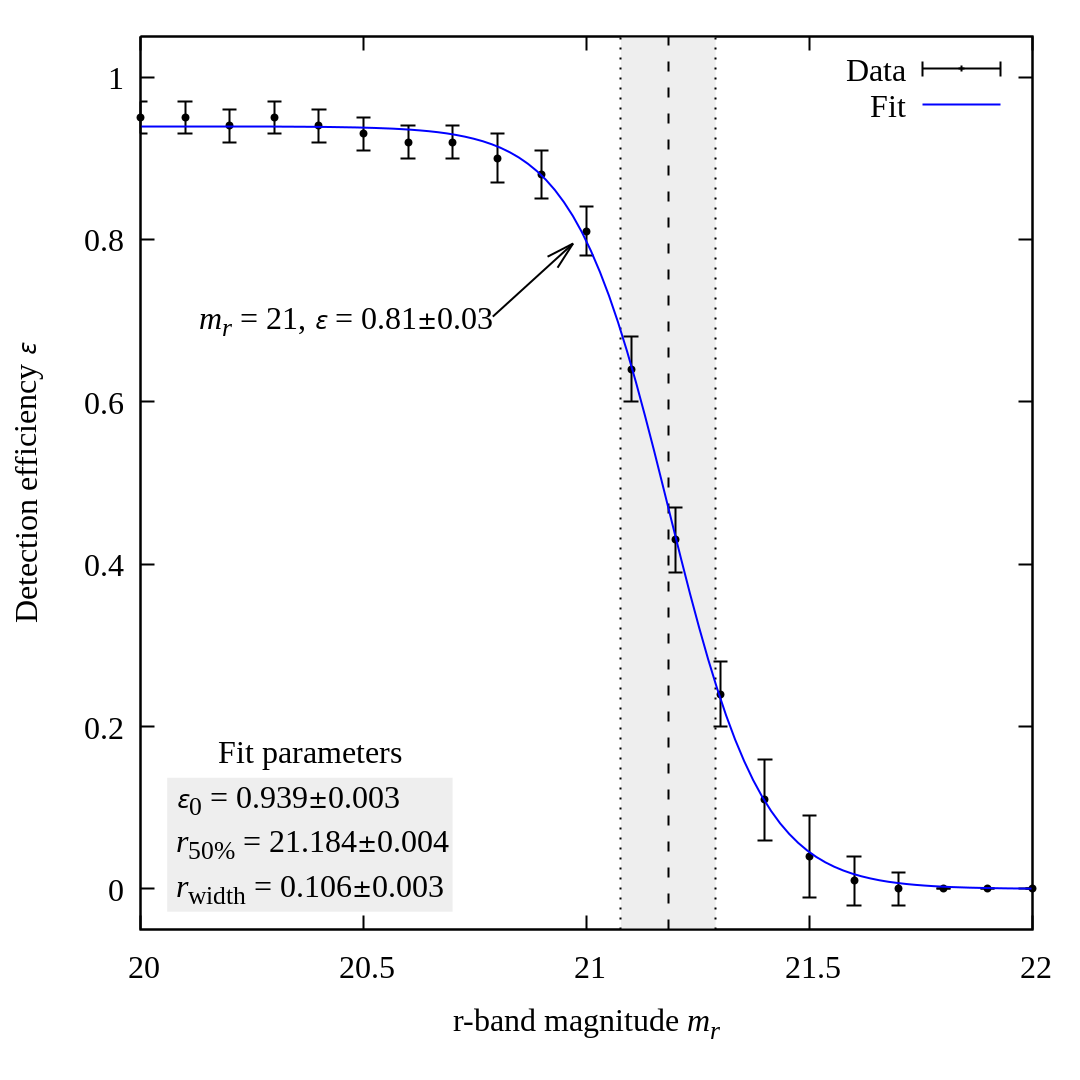}
\caption{\label{FIG:EFFICIENCY}Our CTIO survey moving object detection efficiency $\epsilon$ with respect to the $r$-band magnitude $m_r$. For each frame in our survey we randomly placed 2,100 fake objects with $m_r$ between 20 and 22 (100 objects per 0.1 step in $m_r$) and apparent motions between 90 and 190 arcseconds per hour moving westward into our data. The uncertainty is calculated from the variance in difference survey frames. We fit the detection efficiency using Eq. \ref{EQ:Efficiency} (blue solid line). The fit parameters are displayed in the gray box in the bottom left corner. The arrow points at our limiting magnitude $m_r=21$, where we recorded $\epsilon=0.81\pm0.03$. The gray area defined as $[r_\mathrm{lim}-r_\mathrm{width},r_\mathrm{lim}-r_\mathrm{width}]$ shows the area where $\epsilon$ drops rapidly. The 50\% drop in $\epsilon$ is marked by the dashed line. }
\end{center}

\end{figure}

In order to test the efficiency of our detection algorithm in our survey, we randomly placed fake objects with r-band magnitudes between $m_{r}=20$ and $m_{r}=22$ and apparent motions between 90 and 190 arcseconds per hour moving westward into our data. Our algorithm was able to detect $>80\%$ of fake objects for $m_{r}\le21$, which we consider our limiting magnitude. \added{The results of our survey detection efficiency test are shown in Fig. \ref{FIG:EFFICIENCY}. We fit the detection efficiency using Eq. (15) from \citet{Jedicke_etal_2016}:
\begin{equation}
    \epsilon(m_r) = \epsilon_0 \left[1+\exp\left(\frac{m_r-r_\mathrm{50\%}}{r_\mathrm{width}} \right) \right]^{-1},
    \label{EQ:Efficiency}
\end{equation}
where $\epsilon_0$ is the maximum survey efficiency (i.e. for bright objects), $r_\mathrm{50\%}$ is the $r$-band magnitude for which the efficiency falls to $0.5\epsilon_0$, and $r_\mathrm{width}$ denotes the interval in $m_r$ for which the efficiency decreases from high to low.}

Because of the short time base ($\sim4$ minutes) between images of the same field on a given night, there were no significant differences in our detection of objects with different apparent motions. We note that more than 50\% of the asteroids simulated in this article appeared at least in two observing fields on two different nights since many of our fields from night to night overlapped. This means that our survey efficiency is likely even higher than 80\% for objects brighter than 21st magnitude in the $r$-band as many of the objects could be discovered independently on two different nights. As a conservative estimate we thus assume our DECam survey observing efficiency $\epsilon=0.8$

In our fields we found, without prior knowledge of their locations, three known near-Earth objects: 1998 MN14, 2004 GA1, and 2011 BO24. No other bodies moving faster than 90 arcseconds per hour were detected in our survey. {Detection of no objects near Venus' orbit puts a strong new constraint on the possible population of stable Venus co-orbitals. However, to interpret this negative detection and compare it to previous negative detections and upper limits, we must construct a model that will allow us to calculate the positions and magnitudes of stable Venus co-orbitals.}

\begin{table}[]
\centering
\begin{tabular}{l|c|c}
                                         \multicolumn{1}{c}{Date}  & RA (hrs) & Dec (deg)  \\ \hline \hline
\parbox[t]{22mm}{\multirow{5}{*}{\rotatebox[origin=c]{45}{\shortstack{September 23rd \\ UTC 23:30}}}} & 14:45:00 & -17.50 \\
                                           & 14:45:00 & -19.50 \\
                                           & 14:49:00 & -17.50 \\
                                           & 14:49:00 & -19.50 \\
                                           & 14:49:00 & -15.50 \\\hline
\parbox[t]{15mm}{\multirow{5}{*}{\rotatebox[origin=c]{45}{\shortstack{September 25th \\ UTC 23:30}}}}  & 14:51:00 & -16.25 \\
                                           & 14:51:00 & -18.25 \\
                                           & 14:51:00 & -20.25 \\
                                           & 14:59:00 & -16.25 \\
                                           & 14:59:00 & -18.25 \\\hline
\parbox[t]{15mm}{\multirow{5}{*}{\rotatebox[origin=c]{45}{\shortstack{September 26th \\ UTC 23:30}}}}  & 14:49:00 & -17.50 \\
                                           & 14:49:00 & -19.50 \\
                                           & 14:49:00 & -15.50 \\
                                           & 14:58:00 & -15.50 \\
                                           & 14:58:00 & -17.50 \\\hline
\parbox[t]{15mm}{\multirow{4}{*}{\rotatebox[origin=c]{45}{\shortstack{September 27th \\ UTC 23:30}}}} & 14:47:00 & -17.50 \\
                                           & 14:47:00 & -19.50 \\
                                           & 14:56:00 & -17.50 \\
                                           & 14:56:00 & -19.50 \\\hline
\parbox[t]{15mm}{\multirow{5}{*}{\rotatebox[origin=c]{45}{\shortstack{September 28th \\ UTC 23:30}}}} & 14:46:00 & -17.50 \\
                                           & 14:46:00 & -19.50 \\
                                           & 14:46:00 & -21.25 \\
                                           & 14:59:00 & -21.25 \\
                                           & 14:59:00 & -19.50
\end{tabular}

\caption{\label{TABLE:CTIO} A summary of our September 23rd, 2019 - September 28th, 2019 survey for Venus co-orbital asteroids. The first column shows the time of the observation, the second column shows the right ascension (RA) in hours:minutes:seconds format, and the third column shows the declination in degrees.} 
\end{table}

\section{The apparent visual magnitudes of Venus co-orbital asteroids.}

  Let us begin constructing our model by considering the albedos and phase functions of various asteroid types.  
We will consider 6 different asteroid types: S, M, E, C, P and D (see Table \ref{TABLE:CLASSES}).
\begin{table*}[]
\centering
\begin{tabular}{ccccccc}
\hline
Type & $N$ & $p_V$           & $G_1$         & $G_2$        & $G_{12}$  & Osk$_\mathrm{2012}$ Type\\ \hline
S    & 30  & $0.22\pm0.05$   & $0.26\pm0.01$ & $0.38\pm0.01$ & $0.27\pm0.01$ & A/E/Q/R/S/V/X\\
M    & 10  & $0.17\pm0.07$   & $0.27\pm0.03$ & $0.35\pm0.01$ & $0.30\pm0.02$ & A/E/Q/R/S/V\\
E    & 6   & $0.45\pm0.07$   & $0.15\pm0.02$ & $0.60\pm0.01$ & $0.07\pm0.02$ & R\\
C    & 34  & $0.061\pm0.017$ & $0.82\pm0.02$ & $0.02\pm0.02$ & $0.86\pm0.03$ & C\\
P    & 7   & $0.042\pm0.008$ & $0.83\pm0.03$ & $0.05\pm0.02$ & $0.84\pm0.03$ & C\\
D    & 6   & $0.049\pm0.022$ & $0.96\pm0.03$ & $0.02\pm0.02$ & $0.93\pm0.03$ & C 
\end{tabular}

\caption{Average $p_V, G_1,G_2$ {and $G_{12}$} parameters for main asteroid classes from \citet{Shevchenko_etal_2016}. {The $G_{12}$ parameter was calculated using Eq. 28 from \citet{Muinonen_etal_2010} \added{as and average of $G_{12}(G_1)$ and $G_{12}(G_2)$. The last column shows the taxonomic types derived from $G_{12}$ values using Table 1 in \citet{Oszkieweicz_etal_2012}. Note, that in \citet{Oszkieweicz_etal_2012} the taxonomy types show larger standard deviations in $G_{12}$ than  those derived from \citet{Shevchenko_etal_2016}, thus multiple taxonomic classes can relate to one value of $G_{12}$.}}} 
\label{TABLE:CLASSES}
\end{table*}
\citet{Muinonen_etal_2010} provides a relation between the diameter $D$, absolute magnitude $H$, and geometric albedo $p_V$
\begin{equation}
    \log_{10}D=3.1236 - 0.2H - 0.5 \log_{10}p_V,
    \label{EQ:log10D}
\end{equation}
where $D$ is in km. Of course, for a fixed geometric albedo $p_V$ the absolute magnitude $H$ changes by $5\log_{10}(D_1/D_2)$, where $D_1/D_2$ is the ratio between two different asteroid sizes. Therefore we can simply assume one asteroid size for all asteroid types and re-scale our results later to represent other asteroid sizes. To complicate the matter a bit, $H$ is defined at opposition, i.e., at zero phase angle $\alpha$. 
Most asteroid observations are made at phase angles below $30^{\circ}$, where for some near-Earth asteroids, observations at phase angles up to $90^\circ$ exist \citep{Shevchenko_etal_2019}. For Venus co-orbitals, our observations cover phase angles beyond $90^\circ$ because Venus is located inside the orbit of Earth. So we are faced with the need to extrapolate phase functions from the literature into a new regime.

Asteroid phase functions 
have been extensively studied \citep[see][for a review]{Muinonen_etal_2010}.
We adopt a functional form with three parameters, $H, G_1, G_2$,  that was originally proposed in \citet{Muinonen_etal_2010} and then revisited by \citet{Penttila_etal_2016}.  We take the average values for $G_1$ and $G_2$ from \citet{Shevchenko_etal_2016} (see Table. \ref{TABLE:CLASSES}). In this system, the reduced visual magnitude (the magnitude of an asteroid at $r\Delta =1$ au), is given by 
\begin{equation}
    V(\alpha)=H-2.5\log_{10}\left[G_1\Phi_1(\alpha)+G_2\Phi_2(\alpha)+(1-G_1-G_2)\Phi_3(\alpha)\right],
    \label{EQ:Valpha1}
\end{equation}
where $\Phi_{1,2,3}$ are basis functions defined by cubic splines \citep[see Appendix A in][]{Penttila_etal_2016} and $\alpha$ is the phase angle. We implemented the \citet{Penttila_etal_2016} equations into a Fortran code that is available at GitHub\footnote{\url{https://github.com/McFly007/AstroWorks/tree/master/Pokorny_etal_2020_VenusCoorbs}}.

Once we have the reduced visual magnitude $V(\alpha)$,  we calculate the apparent visual magnitude $m_V$ by taking into account the distances between the object, the observer and the Sun as follows:
\begin{equation}
    V(\alpha)=m_V - 5 \log(r\Delta),
    \label{EQ:Valpha2}
\end{equation}
where 
the heliocentric distance of the asteroid, $r$, and the distance of the asteroid from the observer on Earth, $\Delta$,  are in au (astronomical units). {Our DECam survey observations are conducted in the Dark-Energy Survey (DES) $r$ filter, which can be converted into the Sloan $r'$ filter using Eq. 15 in \citet{Drlica_etal_2018}; $r=r'-0.102(g'-r')+0.02$. Because the absolute magnitude, $H$, is defined with respect to the Johnson's V filter, we need a conversion factor between these filters. We use the conversion factor $V=r'+0.45(g'-r')-0.03$ from \citep{Smith_etal_2002} and the $(g'-r')=0.58\pm0.12$ main-belt asteroid color distribution derived from Figure 8 in \citet{Ivezic_etal_2001}. This leads to a simple conversion factor between Johnson's V, Sloan $r'$ and DES $r$ filters
\begin{equation}
    m_{r} = m_{r'}-(0.039\pm0.012)=m_V-(0.270\pm0.066).
    \label{EQ:VtoR}
\end{equation}

Note, that the uncertainty in Eq. \ref{EQ:VtoR} is much smaller than that introduced by  $p_V$, and thus can be omitted. We can combine  Eqs. (\ref{EQ:log10D}--\ref{EQ:VtoR})} with distributions of heliocentric orbital vectors obtained from \citet{Pokorny_Kuchner_2019} described below to simulate the visual {and $r$-band} magnitudes of the Venus co-orbitals. {Since there are currently no known stable Venus co-orbital asteroids, for simplicity we assume that all asteroids are perfect uniform spheres that do not experience variations of their visual magnitudes due to their surface properties, shape and rotation \citep[see e.g.,][]{Zappala_etal_1990}.}\added{
\citet{Masiero_etal_2009} showed that for 80\% of their main-belt asteroid sample the rotational variability accounted for $<0.4$ change in magnitude and was strongly correlated with the asteroid shape. Modeling of asteroid observations with large phase angles was discussed in \citet{Zavodny_etal_2008}.}

\section{Predictions for the stable component of Venus's co-orbital asteroid population}

All asteroid population models in this manuscript are constructed using orbital state vectors of 862 Venus co-orbital asteroids determined to be stable for 4.5 Gyr by \citet{Pokorny_Kuchner_2019}; we will call this population PK19. For our first test, we integrated PK19 orbits together with all planets for {11.6} years with 3.6525 day time steps using the \texttt{SWIFT\_RMVS4} numerical integrator \citep{Levison_Duncan_2013}. For the planetary configuration, we used initial position and velocity vectors reflecting the state of the planets on September 21st, 2019 at 00:00 UTC, based on JPL's HORIZONS system\footnote{\url{https://ssd.jpl.nasa.gov/horizons.cgi}}. {At every integration time step (including the initial time step, $t = 0$) we recorded the orbital state vectors of all asteroids in our simulation, which yielded $1160\times862=1,000,782$ individual records.}

We assume, as a reference point, that all the asteroids have diameters of 500 m and we examine how the apparent visual magnitudes, $m_V$, of the population are affected by different assumptions of asteroid size and type, choosing among the six different asteroid types described above.

\begin{figure}
\includegraphics[width=\textwidth]{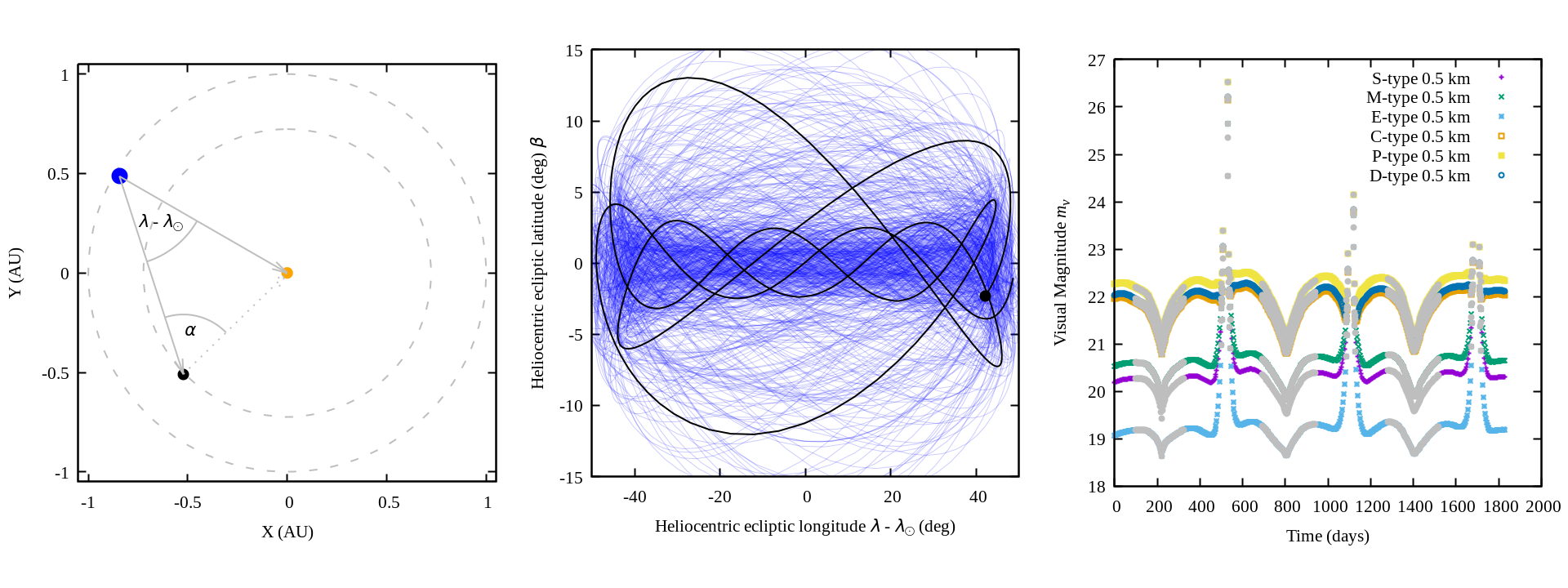}
\caption{Example of how one Venus co-orbital evolves in our model. 
Left panel: Face-on view on the inner solar system. Dashed gray lines represent the mean orbits of Earth and Venus. The blue dot represents the position of Earth, the black dot the position of a Venus co-orbital, and the gray arrows represent the heliocentric and Earth-asteroid vectors. The phase angle $\alpha$ and the heliocentric ecliptic longitude $\lambda-\lambda_\odot$ are shown as grey arcs. 
Middle panel: The {gray} line shows the path of {the} Venus co-orbital during the 5 years of the integration as viewed from Earth in heliocentric ecliptic coordinates (the Sun is at the origin). The black dot shows the position of this particle at the end of the integration. {Blue lines show the orbital evolution of another 19 Venus co-orbitals to better describe the range of the co-orbital motion}.  Right panel: Changes of $m_V$ for the first 5 years of the numerical integration. Color coding shows different asteroidal types based on the \citet{Shevchenko_etal_2016} parameters. The gray areas represent areas where the heliocentric ecliptic longitude of the asteroid $|\lambda - \lambda_\odot|<30^\circ$.}
\label{FIG:ORB_DIAG}
\end{figure}

Figure \ref{FIG:ORB_DIAG} shows the evolution of heliocentric ecliptic longitude $\lambda - \lambda_\odot$ and latitude $\beta$, and apparent visual magnitude $m_V$ for one co-orbital asteroid from our sample. The quantity $\lambda$ is the ecliptic longitude of the asteroid, and $\lambda_\odot$ is the ecliptic longitude of the Sun. 
The heliocentric ecliptic longitude and latitude are given by:
\begin{eqnarray}
    \lambda -\lambda_\odot &=& \mathrm{atan2}(\vec{E_\bot}\cdot \vec{\Delta}, \vec{E_\mathrm{\odot}}\cdot\vec{\Delta} ),\label{EQ:atan2} \\ 
    \beta &=& \mathrm{asin} ( \vec{E_z} \cdot \vec{\Delta} / || \vec{\Delta} || ),
\end{eqnarray}
where $\vec{\Delta}=\vec{r}-\vec{r_\mathrm{obs}}$ is the relative vector between the heliocentric vector of the asteroid $\vec{r}$ and the observer $\vec{r_\mathrm{obs}}$, and $(\vec{E_\odot}, \vec{E_\bot}, \vec{E_z})$ is the orthonormal vector base defined as follows
\begin{eqnarray}
\vec{E_\odot} & = & -\vec{r_\mathrm{obs}}/||\vec{r_\mathrm{obs}}||, \\
\vec{E_z} & = & (\vec{V_\mathrm{obs}} \times \vec{E_\odot})/||\vec{V_\mathrm{obs}}||, \\
\vec{E_\bot} & = & \vec{E_\odot}\times\vec{E_z},
\end{eqnarray}
where $\vec{V_\mathrm{obs}}$ is the velocity vector of the observer. In Eq. \ref{EQ:atan2} we use the 2-argument arctangent, atan2$(y,x)$, a function that allows an unambiguous conversion from Cartesian to polar coordinates, since it returns angles between $0$ and $2\pi$.  

The observer-target-Sun geometry, illustrated in the middle panel of Fig. \ref{FIG:ORB_DIAG}, prescribes that the asteroid spends most of its time in the vicinity of the Sun between $-40^\circ<\lambda -\lambda_\odot<40^\circ$, as seen in the middle panel of Fig. \ref{FIG:ORB_DIAG}.
In the right panel of Fig. \ref{FIG:ORB_DIAG} we assign this one asteroid six different types based on the \citet{Shevchenko_etal_2016} parameters and display how its visual magnitude changes during its orbit, assuming a diameter of 0.5 km.  If we assume the asteroid is an E, S, or M-type, this asteroid can be brighter than $m_V = 21.0$, making it detectable to multiple ground-based telescopes during twilight surveys.

\begin{figure}
\includegraphics[width=\textwidth]{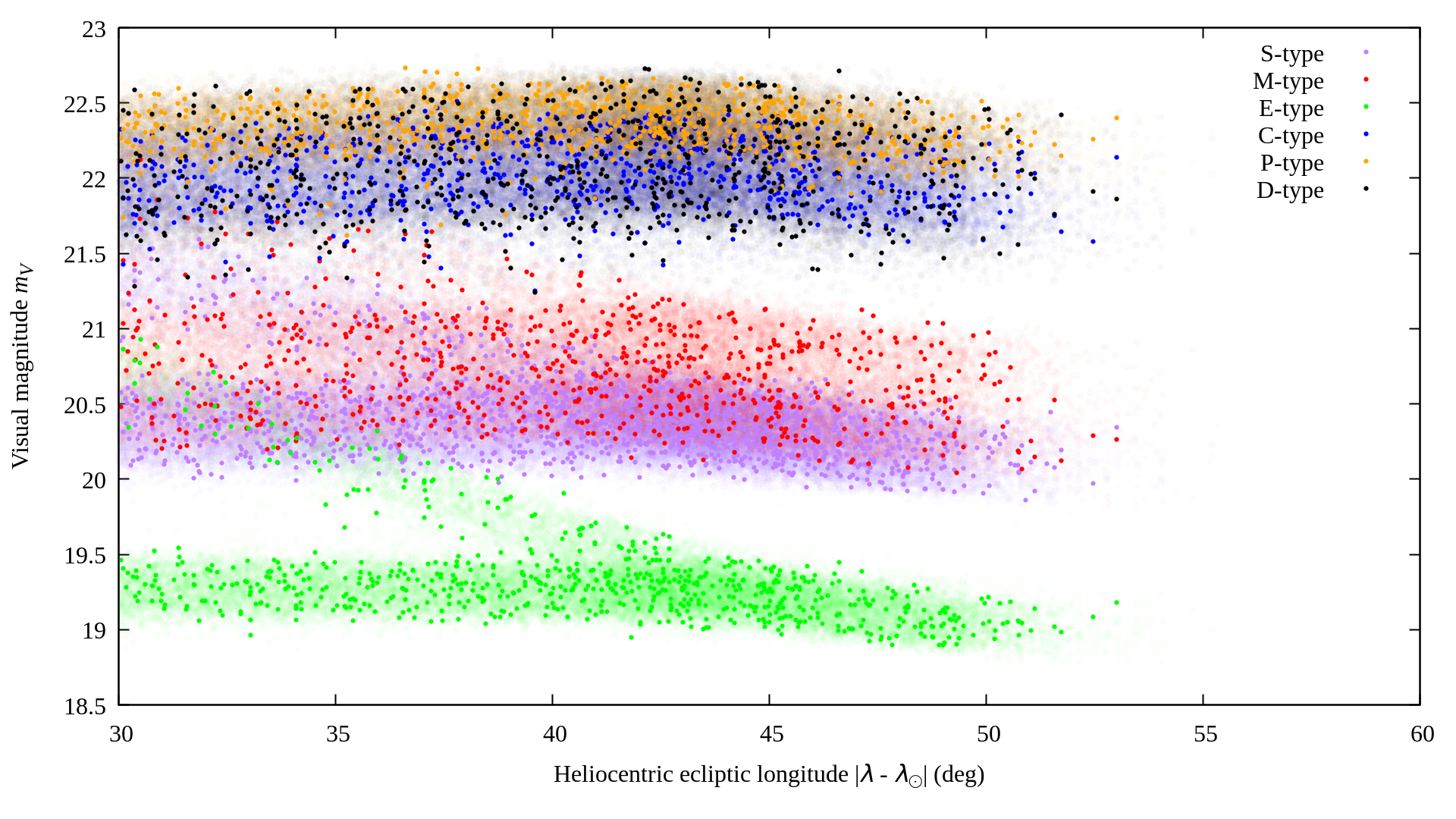}
\caption{Apparent visual magnitudes ($m_V$) and heliocentric ecliptic longitudes for stable Venus co-orbitals in our model, given 6 different asteroid classes. We assumed $D=0.5$ km, $p_V$ {G1, and G2 values with their uncertainties} from Table \ref{TABLE:CLASSES}. {The effect of uncertainties in $p_V, G_1, G_2$ is shown as a light colored background. }}
\label{FIG:LONGITUDE0.5}
\end{figure}

To obtain an overview of the magnitude variations, we plotted the positions of all co-orbital asteroids in our sample tracked for 10 years with 3.6525 days time steps in Figure \ref{FIG:LONGITUDE0.5}. Like the single asteroid shown in Fig. \ref{FIG:ORB_DIAG}, the S, M, and E-type asteroids (purple, red, and green clouds of points) are easier to detect even at larger heliocentric latitudes, with asteroids grazing $\lambda -\lambda_\odot=55^\circ$. Fig. \ref{FIG:ORB_DIAG} shows that Venus co-orbitals can reach higher heliocentric ecliptic latitudes, however, they spend most of their time near the ecliptic. 

\begin{figure}
\includegraphics[width=\textwidth]{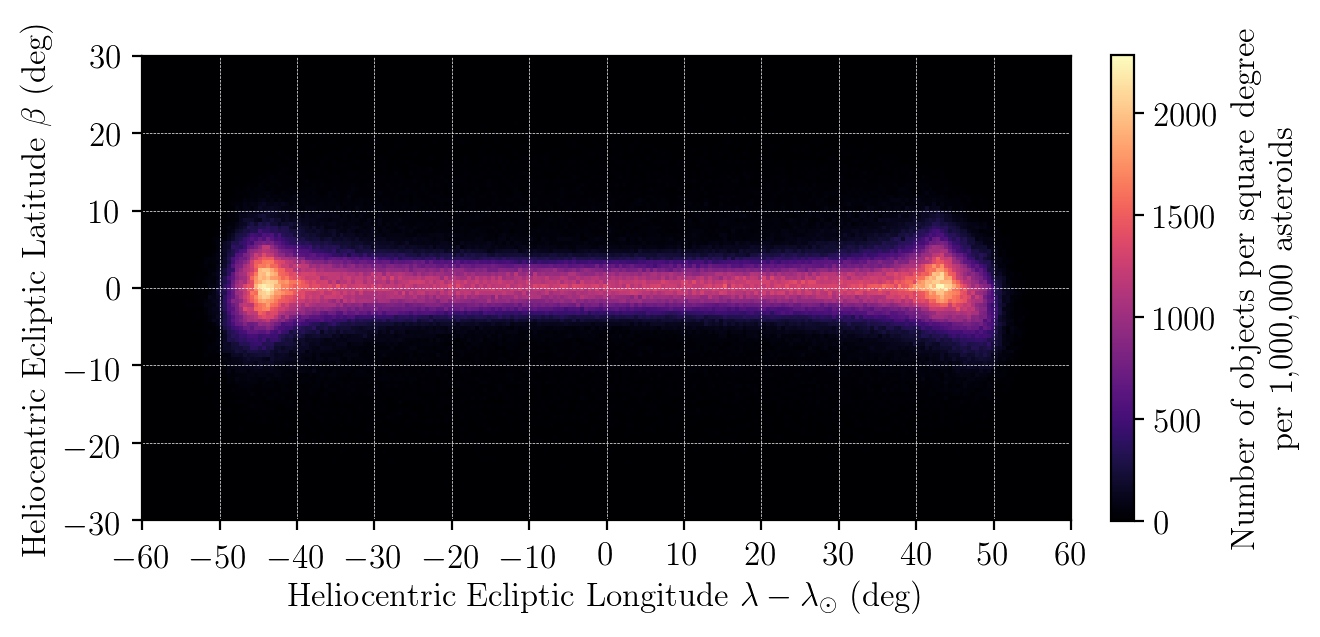}
\caption{Density of asteroids per square degree in heliocentric ecliptic coordinates considering a population of 1,000,000 asteroids co-orbiting with Venus. For this simulation, the orbital motion of Venus has not been subtracted from the positions of the asteroids, so this image represents a time-averaged distribution, rather than a snapshot of a particular moment.}
\label{FIG:ANGLES}
\end{figure}

Fig. \ref{FIG:ANGLES} shows the number of objects per square degree (in $\lambda-\lambda_\odot, \beta$ phase space) for a synthetic population of {1,000,782} Venus co-orbitals, where the location of Venus is assumed to be random (i.e. Fig. \ref{FIG:ANGLES} represents an average number density). 
While Fig. \ref{FIG:LONGITUDE0.5} reveals that orbits of some Venus co-orbitals extend to $\lambda-\lambda_\odot=\pm55^\circ$, Fig. \ref{FIG:ANGLES} shows that the number density of co-orbitals in these regions is fairly small. It also reveals the population of asteroids to be concentrated within a few degrees of the ecliptic. The number density peaks at two different longitudes: around $\lambda-\lambda_\odot=\pm45^\circ$.  At the peaks, the number density is roughly 2,200 asteroids per square degree assuming the total population of 1,000,000 asteroids. These maxima are slightly inward of the extremes of Venus's orbit in our simulation, at $\lambda_V-\lambda_\odot=\pm46.3^\circ$).

\section{Upper limit for the Venus co-orbital population from Sheppard \& Trujillo (2009) observations}

Prior to our CTIO observing campaign described above, \citet{SHEPPARD_TRUJILLO_2009} (henceforth ST2009) performed an observational study of possible asteroids/moons in Venus's vicinity using the Baade-Magellan 6.5 m telescope and IMACS wide-field CCD imager at Las Campanas observatory in Chile.  They took five snapshots of Venus and its vicinity on October 7th, 2005 and attained a maximum limiting apparent R-band magnitude of $m_R=20.4$, corresponding to objects of approximately $D=0.6$ km (assuming albedo $p_V=0.1$).
\citet{SHEPPARD_TRUJILLO_2009} did not detect any asteroids (or moons). We can now use this fact to constrain the co-orbital population with help from our models.

All observations in ST2009 were performed within a span of 30 minutes, so we can consider them as one single observation for our purpose.
However, the ST2009 observation was complicated by the scattered light from Venus, which lead to a variable limiting $m_{R}$ across their field-of-view.  For simplicity, we assume their average limiting $R$-band magnitude to be $m_{R}=19.8$ and the corresponding limiting V-band magnitude to be $m_V=19.8$ mag (we use the conversion from R to V-band magnitudes from Section 7.1 in \citet{Waszcak_etal_2015}: $R=V+(0.00\pm0.10)$.).
Table \ref{TABLE:DIAMS} shows the asteroid diameter $D_\mathrm{lim}$ corresponding to this limiting magnitude ($m_R=19.8$) for all six asteroid types considered here. 
The diameter difference between the smallest E-type asteroid and largest P-type asteroid with that magnitude is approximately a factor of 4.12, representing a factor of 70 in mass. 

\begin{table}[]
\centering
\begin{tabular}{c||c|cc|cc}
Type & $m_V (D=0.5\mathrm{~km})$ & $D_\mathrm{lim} (m_R=19.8)$  & $H{(m_R=19.8)}$ & $D_\mathrm{lim} (m_{r}=21.0)$ & $H(m_{r}=21.0)$  \\ \hline
S	& $20.33 \pm 0.19 $	& $0.638 \pm 0.063 $	& $18.24 \pm 0.19 $	& $0.416 \pm 0.036 $	& $19.17 \pm 0.19 $\\
M	& $20.67 \pm 0.28 $	& $0.746 \pm 0.125 $	& $18.31 \pm 0.28 $	& $0.486 \pm 0.072 $	& $19.24 \pm 0.28 $\\
E	& $19.22 \pm 0.17 $	& $0.383 \pm 0.030 $	& $18.57 \pm 0.17 $	& $0.249 \pm 0.017 $	& $19.50 \pm 0.17 $\\
C	& $22.04 \pm 0.22 $	& $1.403 \pm 0.168 $	& $17.92 \pm 0.22 $	& $0.914 \pm 0.097 $	& $18.85 \pm 0.22 $\\
P	& $22.35 \pm 0.17 $	& $1.618 \pm 0.142 $	& $18.02 \pm 0.17 $	& $1.054 \pm 0.081 $	& $18.95 \pm 0.17 $\\
D	& $22.11 \pm 0.31 $	& $1.449 \pm 0.267 $	& $18.09 \pm 0.31 $	& $0.944 \pm 0.154 $	& $19.02 \pm 0.31 $
\end{tabular}

\caption{\label{TABLE:DIAMS}Average apparent visual magnitudes $m_V$ of Venus co-orbitals assuming six different asteroid types for heliocentric ecliptic longitudes $40^\circ\le|\lambda-\lambda_\odot|\le55^\circ$. The first column shows the asteroid type, the second column shows an average $m_V$ assuming an asteroid diameter $D=0.5$ km. The third and {fifth} columns show the limiting diameters (i.e. the smallest observable diameter) for the ST2009 observation  with the limiting R-band magnitude $m_R=19.8$ mags, and DECam limiting $r$-band magnitude $m_{r}=21.0$, respectively. {The fourth and sixth columns show the absolute magnitudes $H$ calculated from $D_\mathrm{lim}$ for $m_R=19.8$ and $m_{r} = 21.0$.}}

\end{table}

The simulation shown in Figure \ref{FIG:ANGLES} does not match the observing geometry of the \citet{SHEPPARD_TRUJILLO_2009} observations; we need a simulation where Venus's position is fixed with respect to the Earth corresponding to  $\alpha=76.3\pm0.5^\circ$, and $\Delta=0.8782\pm0.01$ au. Such geometry allows for two positions of Venus with respect to Earth, trailing and leading. The leading hemisphere is defined as the portion of the sky centered around the observer's (Earth's) velocity vector. At the time of the ST2009 observation Venus was trailing behind Earth, $\lambda-\lambda_\odot=-45^\circ$.

Such a specific geometry is significantly different from our simulation summarized in Fig. \ref{FIG:ANGLES}. To get enough statistics we ran a new simulation of  $N=862$ PK2019 asteroids and created a snapshot of their positions every time the Earth-Venus geometry resembled that of the \citet{SHEPPARD_TRUJILLO_2009} observations, i.e., $\alpha=76.3\pm0.5^\circ$, and $\Delta=0.8782\pm0.01$ au. We integrated the orbits of all asteroids and planets for $5\times10^6$ days (13690 years) with a 2 day time step and we obtained a total sample of 1,755,034 records of co-orbitals' orbital elements, 765,457 for the configuration when Venus is trailing behind Earth, and 989,577 for the leading configuration. 

\begin{figure}
\includegraphics[width=\textwidth]{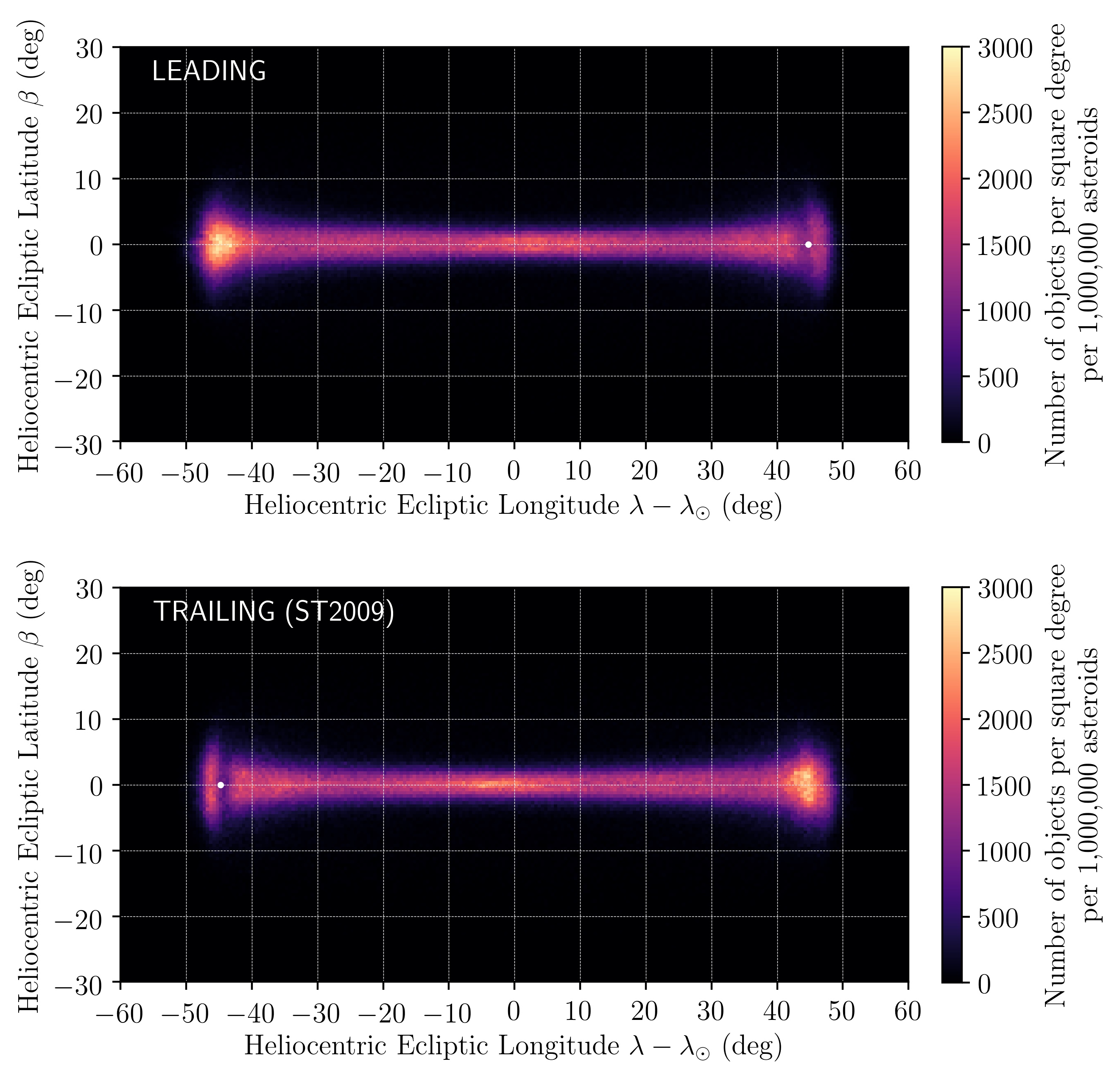}
\caption{The same as Figure \ref{FIG:ANGLES} but now for a specific observing configuration similar to \citet{SHEPPARD_TRUJILLO_2009}. The figure shows two different Earth-Venus configurations: Top panel shows Venus leading, while the bottom panel shows Venus trailing, which is the configuration from \citet{SHEPPARD_TRUJILLO_2009}. For both configurations, the area around Venus (white dot close to $\lambda-\lambda_\odot = \pm 44.85^\circ$) is significantly depleted in the number of asteroids compared to the anti-Venus point. The best place to search for Venus co-orbitals is the limb of Venus's orbit, as long as Venus itself is elsewhere.}
\label{FIG:ANGLES_TS2009}
\end{figure}

Figure \ref{FIG:ANGLES_TS2009} shows the results of our simulation with the ST2009 observing geometry, where the top panel shows the Venus leading case, whereas the bottom panel shows the trailing case (white dot represents the location of Venus). When compared with Fig. \ref{FIG:ANGLES}, it is evident that for both the leading and trailing case, Venus is located at a local minimum of the asteroid number density, while the anti-Venus point presents a global maximum with a {factor of} 3--4 higher number density. This phenomenon reflects the fact that Venus creates a particle-depleted cavity in its vicinity for co-orbiting meteoroids \citep{Pokorny_Kuchner_2019}. The depletion is a consequence of the three-body dynamics and the shape of the zero-velocity curves \citep[see e.g. Chapter 3 in][]{Dermott_etal_1994}.

\begin{figure}
\includegraphics[width=\textwidth]{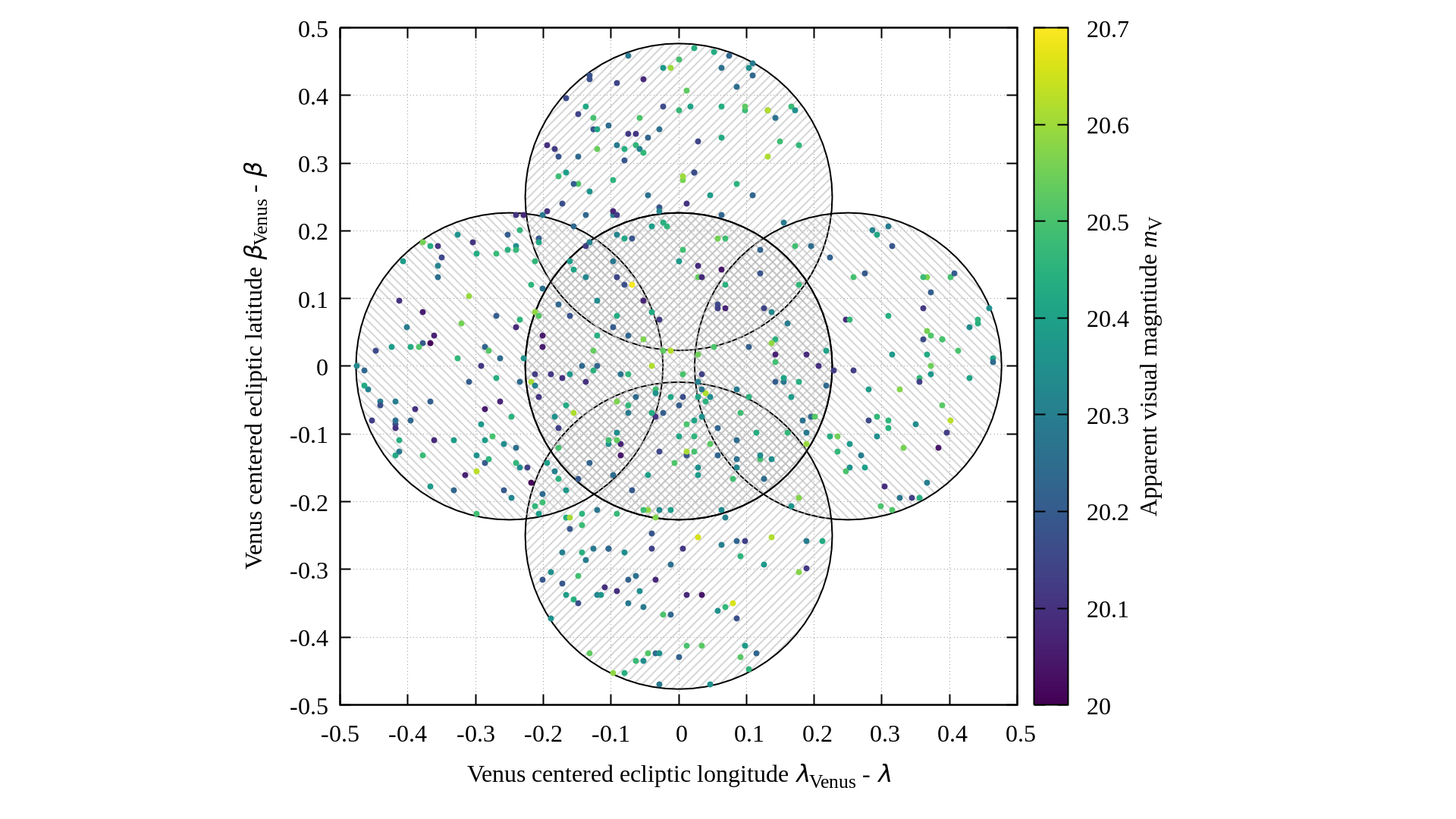}
\caption{\citet{SHEPPARD_TRUJILLO_2009} observation regions (shaded areas) and 478 synthetic Venus's co-orbitals color coded by their apparent visual magnitude $m_V$ assuming that co-orbitals are S-types with $D=0.5$ km. The total number of asteroids in this simulation was 765,457. Using bootstrapping we estimate that \citet{SHEPPARD_TRUJILLO_2009} could have detected 1 out of every 1565 stable Venus co-orbital asteroids, assuming this size and type.}
\label{FIG:ST2009}
\end{figure}

Now, let us focus on the bottom (trailing) panel of Fig. \ref{FIG:ANGLES_TS2009}, the exact geometry of the ST2009 observation.
Figure \ref{FIG:ST2009} shows 478 synthetic asteroids that fell within the regions observed by \citet{SHEPPARD_TRUJILLO_2009} (shaded areas). Assuming all asteroids to be S-types with $D=0.5$ km, the apparent visual magnitude dispersion is quite small: $m_V=20.33\pm0.15$ mags. When we compare all six asteroid types assumed here, the uncertainty of $m_V$ is in the range of {0.12--0.29} mags. Comparing the number of simulated asteroids within the overall field-of-view depicted in Fig. \ref{FIG:ST2009} to the total number of simulated asteroids, we derive an observing probability $\mathcal{P_\mathrm{T}}=478/765457=0.0624\%$, i.e., 1 in 1600 asteroids appears in the field-of-view. 
This probability estimate has an uncertainty from Poisson noise, which we calculate to be $1/\sqrt{478}=0.0457$, i.e $\mathcal{P_\mathrm{T}}=478/765457=0.0624\% \pm 0.0029\%$.  We confirm this uncertainty with a bootstrapping calculation detailed below.

To better understand the statistical mean $\mu$ and standard deviation $\sigma$ of the $\mathcal{P}_\mathrm{T}$ distribution, we performed bootstrapping of the entire simulated asteroid population 10,000 times; i.e., we randomly picked asteroids from the entire population until we obtained the original number allowing sampling with replacement (allowing for double counting). Bootstrapping resulted in a normal distribution of $\mathcal{P}_\mathrm{T}$ drawn from our simulation sample, with a resulting observing probability of $\mathcal{P_{\mathrm{T}_\mathrm{Boot}}}=0.0639\% \pm 0.0029\%$, which is slightly higher than inferred from the simulation, while the uncertainty matches the Poisson noise.

When we relax the requirement that Venus is trailing behind Earth while keeping the same observation geometry (i.e., the same values for $\alpha$ and $\Delta$). For the configuration when Venus is leading we recorded 658 asteroids in the potential field-of-view out of the total sample of 989,577 asteroids. This translates into the leading configuration observing probability (asteroid appeared in the ST2009 field-of-view) of $\mathcal{P}_\mathrm{L}=658/989577=0.066\%$, while bootstrapping provides again a slightly higher probability $\mathcal{P_{\mathrm{L}_\mathrm{Boot}}}=0.0687\% \pm 0.0026\%$. When we sum both the trailing and leading configurations we get  $\mathcal{P}_\mathrm{T+L}=1136/1755034=0.064\%$ and a bootstrapped value of $\mathcal{P_{\mathrm{T+L}_\mathrm{Boot}}}=0.0667\% \pm 0.0019\%$. Bootstrapping also shows that the amount of co-orbital asteroids in our simulation is satisfactory because for all three options $\sigma<0.05\mu$.

Let us assume that at the time of the \citet{SHEPPARD_TRUJILLO_2009} observation there were $\mathcal{N}_\mathrm{tot}$ Venus co-orbitals that would appear in their field-of-view brighter than their limiting magnitude.  From our simulation we calculate that each such asteroid had a probability of appearing in the field-of-view of $\mathcal{P_{\mathrm{T}_\mathrm{Boot}}}=0.0639\% \pm 0.0029\%$. 
Then, the probability of detecting zero asteroids, $\mathcal{P_\mathrm{miss}}$, decreases with the increasing number of detectable asteroids, $N_\mathrm{tot}$, as follows:
\begin{equation}
    \mathcal{P_\mathrm{miss}} = (1-\mathcal{P_{\mathrm{T}_\mathrm{Boot}}})^{\mathcal{N}_\mathrm{tot}}.
    \label{EQ:PMISS}
\end{equation}
In return, we can rewrite Eq. \ref{EQ:PMISS} into a probability of detecting at least one asteroid $\mathcal{P}_\mathrm{detect}=1-\mathcal{P}_\mathrm{miss}$. 
Table \ref{TABLE:PROBS} shows five different values of $\mathcal{N}_\mathrm{tot}$, and the corresponding values of $\mathcal{P}_\mathrm{miss}$ and $\mathcal{P}_\mathrm{detect}$.  
For example, we find that the population of co-orbitals brighter than $m_R=19.8$ would need to be $\mathcal{N}_\mathrm{tot}=1084\pm49$ or greater for \citet{SHEPPARD_TRUJILLO_2009} to have had at least a 50\% chance of detecting one.

\begin{table}[]
\centering
\begin{tabular}{ccc}
\hline
$\mathcal{P_\mathrm{miss}}$ & $N_\mathrm{tot}$ &  $\mathcal{P_\mathrm{detect}}$ \\ \hline
0.90    & $164\pm7$   &0.10\\
0.75    & $450\pm20$  &0.25\\
0.50    & $1084\pm49$ &0.50\\
0.25    & $2169\pm98$ &0.75\\
0.10    & $3602\pm164$& 0.90
\end{tabular}
\caption{The upper limit for the total number of co-orbital asteroids larger than the limiting magnitude ($m_R=19.8$) from \citet{SHEPPARD_TRUJILLO_2009} for different values of $\mathcal{P_\mathrm{miss}}$ {and $\mathcal{P_\mathrm{detect}}$}. The limiting magnitude corresponds to 400-1500 m asteroid in diameter depending on their spectral type (see Table \ref{TABLE:DIAMS}).} 
\label{TABLE:PROBS}
\end{table}

\section{Upper limit for the Venus co-orbital population from DECam}

Now let us turn to our new non-detection derived from our DECam instrument survey. In order to predict the positions of Venus co-orbitals at the exact time of our DECam observations, we evolved this same sample of 862 stable co-orbitals and Venus until the ecliptic longitude of Venus in the simulation matched that of Venus on September 23rd, 2019 at UTC 23:30. Due to slightly different orbital elements of planets from the 4.5 Gyr simulation in \citet{Pokorny_Kuchner_2019} and the current planetary orbital elements, it was not possible to use the 4.5 Gyr simulation to obtain precise predictions for locations of Venus co-orbitals in the sky. To compensate for this difference, we extracted the position and velocity vectors of the 862 surviving asteroids and placed them in the planetary configuration of September 23rd, 2019 at UTC 23:30 (i.e., the one where the ecliptic longitude of Venus matches the simulation). Then we numerically evolved all simulated asteroids with a 2 hour time-step for 1 million days (2738 years) and recorded their RA and Dec every time they matched the observation geometry (i.e. the positions of Venus and Earth were the same as the time of observation). In total we obtained 439,620 unique records of asteroids co-orbiting Venus matching our observing conditions. Then we applied the DECam field-of-view centered at the coordinates given in Table \ref{TABLE:CTIO} for each day and counted all unique asteroids in our sample, i.e. we do not double count any asteroids that could have been detected on multiple occasions. {We also excluded all asteroids with RA and Dec drift below 90 arcseconds per hour.} This analysis led to {21,185} simulated asteroids that were in the field-of-view of our DECam survey, which is {$4.82\%$} of the simulated population. In other words, we find $\mathcal{P}=21185/439,620 = 0.0482 \pm 0.003$. {The uncertainty of $\mathcal{P}$ can be neglected.}

\begin{figure}
\includegraphics[width=\textwidth]{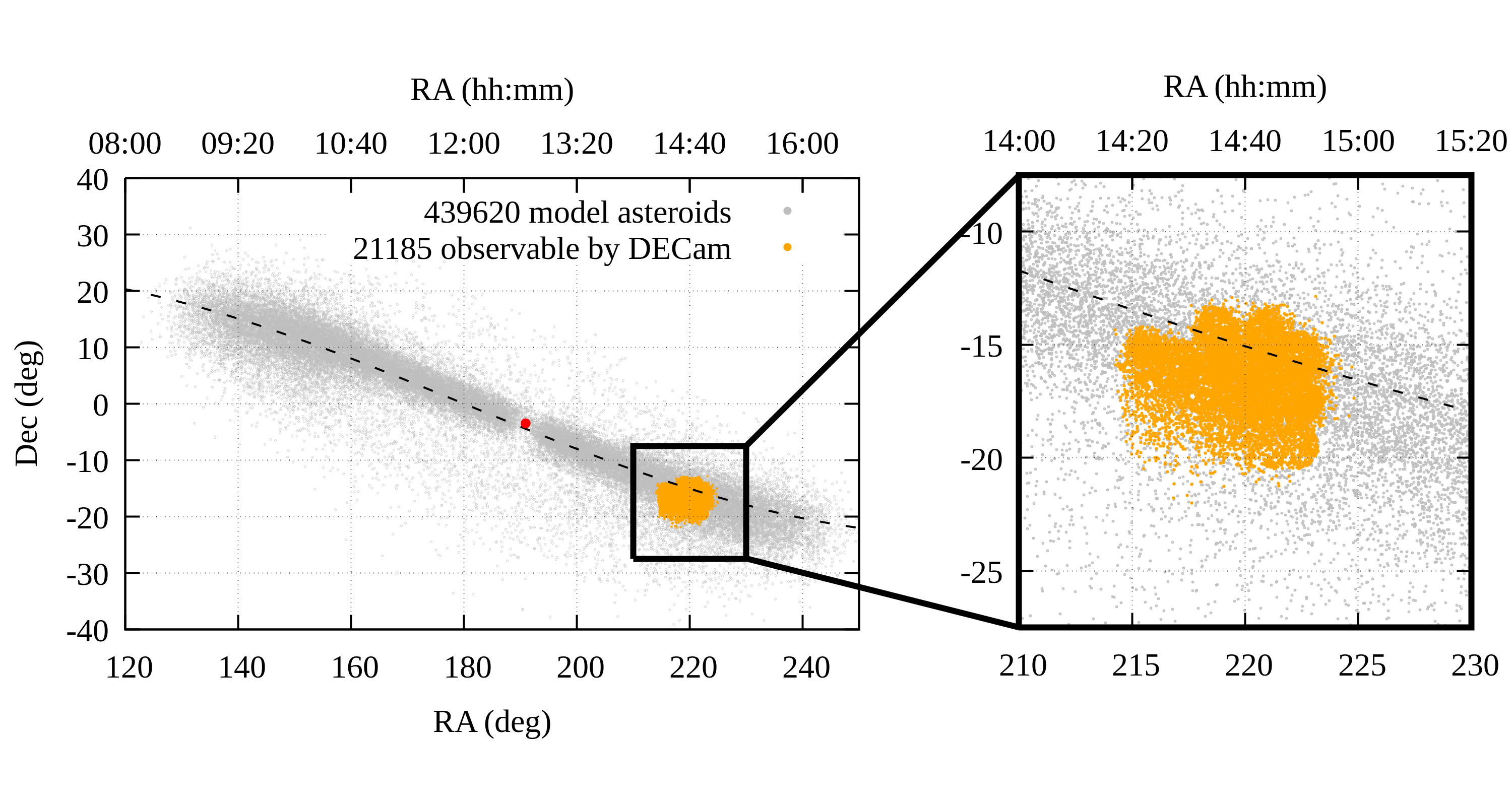}
\caption{Left: Distribution of 439,620 simulated Venus co-orbitals as they would appear on September 23rd at 23:30 UTC (gray circles) and 21,185 asteroids that were in the field-of-view of DECam during our 5 day survey (orange circles). The black dashed line represents the ecliptic, the red dot is the position of Venus on September 23rd at 23:30 UTC. Right: The same as the left panel but now zoomed in to $210^\circ<\mathrm{RA}<230^\circ$ and $-25^\circ<\mathrm{Dec}<-5^\circ$. In total, our DECam survey should have observed approximately $4.82\%$ of the entire population.}
\label{FIG:CTIO}
\end{figure}

Figure \ref{FIG:CTIO} shows the distribution of the simulated Venus co-orbitals, and the portion of the simulated population that was in the field-of-view of our survey. The simulated population lies near the ecliptic (black dashed line), where we predict the densest portion of the leading population. Indeed, we planned our DECam observations with this prediction in hand, purposefully probing this high density region around $\mathrm{RA}=220^\circ$ and $\mathrm{Dec}=-17.5^\circ$. A similar predicted high density region lurks at a positive declination ($\mathrm{RA}=160^\circ$ and $\mathrm{Dec}=+12.5^\circ$), but this location was not observable from our location. \citet{Pokorny_Kuchner_2019} showed that the co-orbital population should be symmetric in the longitudinal direction, so we have no reason to think that observing only one high density region significantly alters the conclusions of our survey.

In previous paragraphs we showed that our DECam survey should have seen on average $\mathcal{P}=4.82\%$ of the Venus co-orbital population, with a total number $\mathcal{N}_\mathrm{tot}$ of co-orbital asteroids that would appear in our field-of-view  brighter than $m_{r}=21$. $\mathcal{N}_\mathrm{tot}$ is not the total number of asteroids co-orbiting with Venus, but rather the total number of asteroids that are observable by our survey, i.e. their $r$-band apparent magnitude is brighter than 21, which translates to asteroid diameters shown in Table \ref{TABLE:DIAMS}. {Taking into account the survey efficiency $\epsilon=0.8$ described in Section 2, and} combining $\mathcal{N}_\mathrm{tot}$ and $\mathcal{P}$ gives an expected number of Venus co-orbitals in our field-of-view $\mathcal{N}=\epsilon \mathcal{N}_\mathrm{tot}\mathcal{P}$. Statistically, observing asteroids at any given time should follow Poisson statistics, with the mean $\lambda_P=\mathcal{N}$, where the probability of occurrence for $k$ events for a given mean $\lambda_P$ is:
\begin{equation}
    P(k,\lambda_P)=\frac{\lambda_P^k e^{-\lambda_P}}{k!} \Rightarrow P(k,\mathcal{N})=\frac{\mathcal{N}^k e^\mathcal{N}}{k!}; \mathcal{N}=\epsilon \mathcal{N}_\mathrm{tot}\mathcal{P}.
    \label{EQ:Poisson}
\end{equation}
Let us assume that in our survey we had a 50\% chance to observe at least one Venus co-orbital and a 50\% chance to observe zero co-orbitals, {i.e, $k=0$}. Using Eq. \ref{EQ:Poisson} we get $P(0,\lambda)=0.5$ for $\mathcal{N}=0.692,\mathcal{N}_\mathrm{tot}=17.98$. A one-$\sigma$ interval $(0.1587<P<0.8413)$ for observing zero asteroids translates into {$(0.16\le\mathcal{N}\le1.85)$ and $(4.48\le\mathcal{N}_\mathrm{tot}\le47.74)$}, while a 3-$\sigma$ interval $(0.0013<P<0.9987)$ represents {$0.001\le\mathcal{N}\le6.65$ and} $0\le\mathcal{N}_\mathrm{tot}\le172$. If we assume, that we were extremely unlucky, zero detected asteroids happens with 1/10000 probability for {$\mathcal{N}=9.21$ or} $\mathcal{N}_\mathrm{tot}=239$, which we can consider as an extreme upper limit. 

\begin{figure}
\includegraphics[width=\textwidth]{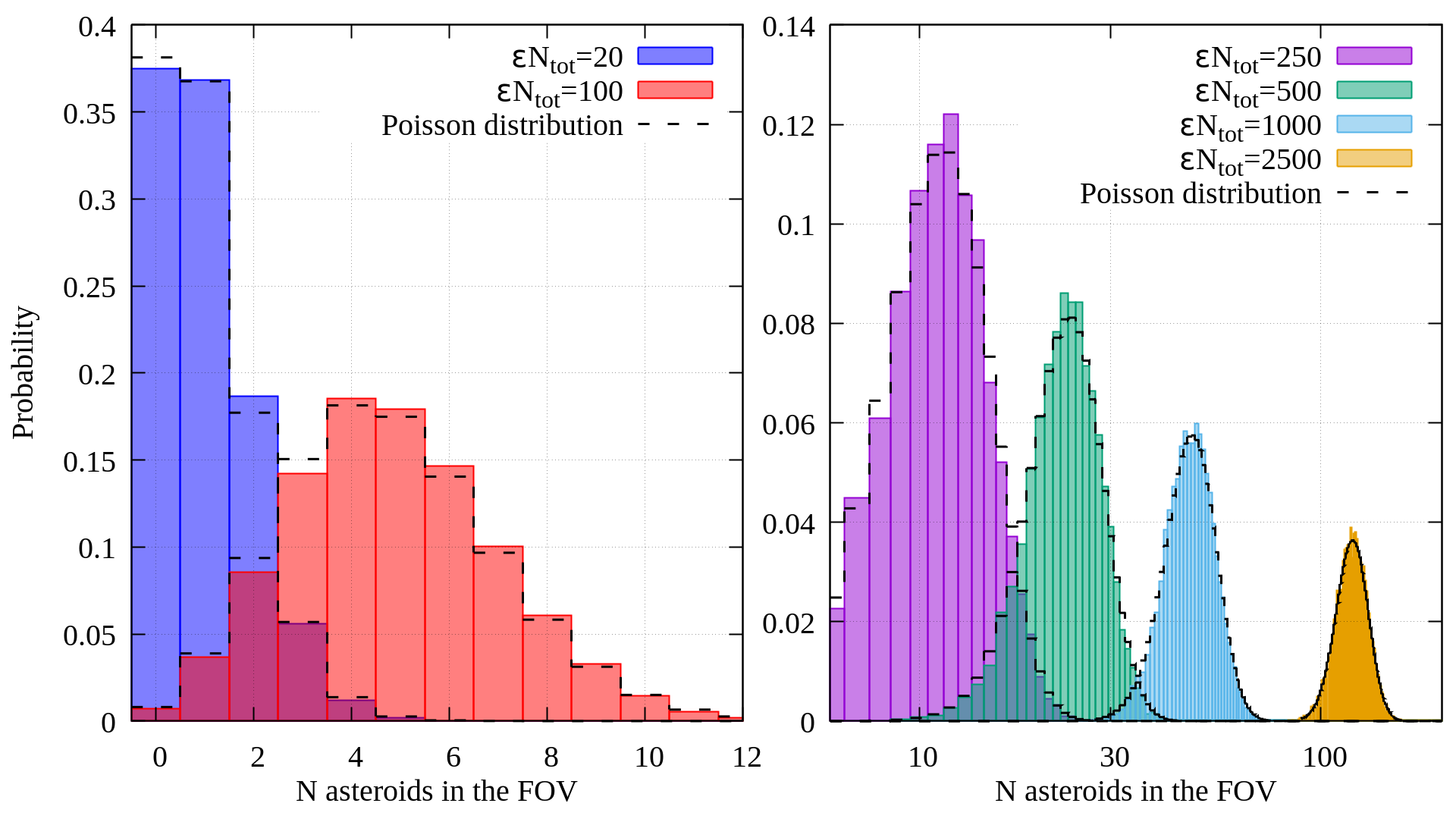}
\caption{Histograms of probability distributions of the number of asteroids in our DECam survey field-of-view based on assumed value of $\epsilon\mathcal{N}_\mathrm{tot}$, the {product of the detection efficiency $\epsilon$ and the} total number of asteroids co-orbiting with Venus brighter than apparent $r$-band magnitude 21{, ${N}_\mathrm{tot}$}. The probability distributions were calculated using 10,000 Monte-Carlo picks from our simulated asteroid population for our DECam survey. \textit{Left:} Blue histogram is corresponding to $\epsilon\mathcal{N}_\mathrm{tot}=20$, red histogram shows $\epsilon\mathcal{N}_\mathrm{tot}=100$, black dashed line is the probability distribution calculated from the Poisson distribution. \textit{Right:} The same as the \textit{left} panel but now for four values of $\epsilon\mathcal{N}_\mathrm{tot}=250;500;1000\mathrm{~and~}2500$. The Poission distribution predicts well the probability distribution regardless of $\mathcal{N}_\mathrm{tot}$.}
\label{FIG:POISSON}
\end{figure}

To check whether our simulated population and its appearance in our field-of-view follows the assumed Poisson distribution we perform bootstrapping of our simulated sample by picking random ($\mathcal{N}_\mathrm{tot} =2500; 1000; 500; 250; 100; \mathrm{~and~} 20$) asteroids from our sample and checking if any of them are within our field-of-view. We repeated this procedure 10,000 times in order to obtain good statistics. Figure \ref{FIG:POISSON} shows results from our Monte-Carlo test, histograms of the number of simulated asteroids in the field-of-view $\mathcal{N}$ for six different values of $\mathcal{N}_\mathrm{tot}$, and their corresponding Poisson distributions. Regardless of the value of $\mathcal{N}_\mathrm{tot}$, Poisson statistics provide a great fit to our Monte-Carlo distributions, which we consider enough to validate our original assumption that the probability of observing $\mathcal{N}$ asteroids can be estimated by Poisson distribution.

Our conclusion from our CTIO DECam survey is that if the synthetic population of asteroids  from \citet{Pokorny_Kuchner_2019} is co-orbiting with Venus, then the upper limit for the total number of objects that would have in our field-of-view apparent magnitudes in the $r$-band $m_{r}<21.0$ is $\mathcal{N}_\mathrm{tot}=18^{+30}_{-14}$. The limiting diameters for the six asteroid classes in our survey can be found in Table \ref{TABLE:DIAMS}. As shown in Fig. \ref{FIG:LONGITUDE0.5} the apparent magnitude can significantly differ for different asteroid types assuming the same asteroid size. {Furthermore, with more surveys providing better estimates and more dynamical models, we will be able to narrow the confidence interval of the number of Venus co-orbital asteroids.} Additionally, due to their proximity to the Sun and long term exposure to temperatures much higher than asteroids in the main-belt, the albedo of these asteroids might be much smaller ($<0.001$) than dark C-types. However, until at least one of the Venus co-orbitals is found, this is pure speculation.

\section{Discussion and Conclusions}
In this manuscript we showed that stable Venus co-orbital asteroids, if they exist, are likely located close to the ecliptic; 32\% are within $1^\circ$, 72\% are within $3^\circ$, and 87\% are within $5^\circ$ of the ecliptic, and within $\pm50^\circ$ of the heliocentric longitude (Figs. \ref{FIG:ANGLES} and \ref{FIG:ANGLES_TS2009}). Co-orbital asteroids experience significant variations of their apparent magnitude depending on the observational geometry of the Earth-asteroid-Sun system (Fig. \ref{FIG:ORB_DIAG}).
Using six different asteroid types we predict that Venus's co-orbitals show little variation in the value of the apparent visual magnitude when observed more at heliocentric longitudes $|\lambda-\lambda_\odot|>40^\circ$ (Fig. \ref{FIG:LONGITUDE0.5}), where the values corresponding to objects with 0.5 km diameter can be found in Table \ref{TABLE:DIAMS}.

Venus co-orbitals show a strong concentration when observed at heliocentric ecliptic longitudes $|\lambda-\lambda_\odot|\sim44^\circ$ (Fig. \ref{FIG:ANGLES}) unless Venus itself happens to be in the vicinity (Fig. \ref{FIG:ANGLES_TS2009}). In this case the asteroid number density sharply drops in the vicinity of Venus because of the shape of the co-orbital asteroidal ring \citep[see Fig. 3C in][for a similar effect on meteoroids]{Pokorny_Kuchner_2019}.  Moreover, the nearby presence of bright Venus tends to increase background noise from scattered light.

We compared our model to a previous search and non-detection by \citet{SHEPPARD_TRUJILLO_2009}, and find that this work surveyed roughly 0.06\% of the co-orbital population down to $m_R =19.8$.  This magnitude limit corresponds to asteroids with diameters larger than $D=0.4$ km, though the limiting diameter is a strong function of the asteroid type, and for less reflecting asteroids (C, P, D-types) the limiting diameter exceeds 1 km.

Our 5-night DECam survey with limiting magnitude of 21 mag in the $r$-band also resulted in no new detections of low eccentricity Venus co-orbitals, yielding a much more stringent constraint on the population. Using our dynamical model for the distribution of Venus co-orbital asteroids and the fields of view of our survey, we find that our survey investigated roughly 5\% of the population of stable Venus co-orbitals down to an apparent $r$-band magnitude of 21. 
 Our new upper limit is supported by the somewhat less sensitive \citet{Ye_etal_2020} survey that detected no stable Venus co-orbital asteroids brighter than 19.5 mag in the $r$-band. We conclude that the upper limit for asteroids brighter than apparent $r$-band magnitude 21 is $\mathcal{N}_\mathrm{tot}=18^{+30}_{-14}$. The $3-\sigma$ interval for $\mathcal{N}_\mathrm{tot}$ is $0-172$ asteroids. Table \ref{TABLE:DIAMS} shows that 21 mag in the $r$-band corresponds to a diameter of {$D=0.416$} km for the most common NEA spectral types, S-types \citep{Devogele_etal_2019}, while for darker C-types the asteroids are around {$D=0.914$} km.
 
 \deleted{\citet{Morais_Morbidelli_2006} predicts $0.14\pm0.03$ Venus co-orbital asteroids with $H<18$ (approximately 0.5 -- 1.6 km in diameter for the six asteroid types assumed here) assuming these co-orbital asteroids are supplied from the main-belt. This predicted number is 1-2 orders of magnitude lower than our upper limit based on the assumption that the Venus co-orbital asteroids are formed and stay in the 1:1 MMR with Venus. The \citet{Granvik_etal_2018} model supports findings of \citet{Morais_Morbidelli_2006} and shows that the transport of asteroids from various reservoirs beyond Venus' orbit to low eccentricity orbits co-orbiting with Venus is inefficient.}  

 \added{\citet{Morais_Morbidelli_2006} predicts $0.14\pm0.03$ Venus co-orbital asteroids with $H<18$ (approximately 0.5 -- 1.6 km in diameter for the six asteroid types assumed here) assuming these co-orbital asteroids are supplied from the main-belt.   The \citet{Granvik_etal_2018} model supports findings of \citet{Morais_Morbidelli_2006} and shows that the transport of asteroids from various reservoirs beyond Venus' orbit to low eccentricity orbits co-orbiting with Venus is inefficient. 
 Both papers predict 1-2 orders of magnitude smaller numbers of stable Venus co-orbitals compared to our upper limit estimates based on our survey's efficiency and non-detection. Our survey would have $<3\%$ chance to detect stable Venus co-orbital asteroids based on the \citet{Morais_Morbidelli_2006} prediction.
 }  

\citet{Pokorny_Kuchner_2019} estimated the total dust mass of Venus's circumsolar dust ring to $1-3 \times 10^{13}$ kg, which is equivalent to a $D=2-3$ km asteroid ground to dust. The dust co-orbiting with Venus is not dynamically stable for more than a million years and so it needs to be replenished continuously or it must originate from a recent event. Using a single-power law for the size-frequency distribution with a differential index $\gamma \in [-3.0,-4.0$], where $\gamma=-3.5$ represents a population in a collisional balance \citet{Dohnanyi_1969}, we predict $<1$ objects with $D>2$ km for S-types and {1--6} objects for C-types to be currently present in the stable co-orbital population of Venus based on our new DECam survey. In other words, we cannot quite reject the hypothesis that a small number of stable Venus co-orbitals exist and supply Venus's circumsolar dust ring.  

\acknowledgments
\noindent Funding: P.P. and M.J.K. were supported by NASA Solar System Workings grant NNH14ZDA001N-SSW and NASA ISFM awards. S.S.S was supported by NASA Planetary Astronomy grant NN15AF446.\\
Author contributions: 
P.P.: Conceptualization, formal analysis, investigation, methodology, software, validation, visualization, writing - original draft, writing - review \& editing \\
M.J.K.: Conceptualization, formal analysis, funding acquisition, investigation, methodology, validation, writing - original draft, writing - review \& editing\\
S.S.S.: Investigation, validation, writing - original draft, writing - review \& editing\\
Competing interests: Authors declare no competing interests.\\
Data and materials availability: All data is available in the main text or the supplementary materials.
\\
\\
This project used data obtained with the Dark Energy Camera (DECam), which was constructed by the Dark Energy Survey (DES) collaborating institutions: Argonne National Lab, University of California Santa Cruz, University of Cambridge, Centro de Investigaciones Energeticas, Medioambientales y Tecnologicas-Madrid, University of Chicago, University College London, DES-Brazil consortium, University of Edinburgh, ETH-Zurich, University of Illinois at Urbana-Champaign,Institut de Ciencies de l’Espai, Institut de Fisica d’Altes Energies, Lawrence Berkeley National Lab, Ludwig-Maximilians Universitat, University of Michigan, National Optical Astronomy Observatory, University of Nottingham, Ohio State University, University of Pennsylvania, University of Portsmouth, SLAC National Lab, Stanford University, University of Sussex, and Texas A\&M University. Funding for DES, including DECam, has been provided by the U.S. Department of Energy, National Science Foundation, Ministry of Education and Science (Spain), Science and Technology Facilities Council (UK), Higher Education Funding Council (England), National Center for Supercomputing Applications, Kavli Institute for Cosmological Physics, Financiadora de Estudos e Projetos, Fundao Carlos Chagas Filho de Amparo a Pesquisa, Conselho Nacional de Desenvolvimento Cientfico e Tecnolgico and the Ministrio da Cincia e Tecnologia (Brazil), the German Research Foundation-sponsored cluster of excellence ”Origin and Structure of the Universe” and the DES collaborating institutions.

\bibliography{papers}

\begin{thebibliography}{}
\expandafter\ifx\csname natexlab\endcsname\relax\def\natexlab#1{#1}\fi

\bibitem[{{Brasser} {et~al.}(2004){Brasser}, {Innanen}, {Connors}, {Veillet},
  {Wiegert}, {Mikkola}, \& {Chodas}}]{Brasser_etal_2004}
{Brasser}, R., {Innanen}, K.~A., {Connors}, M., {et~al.} 2004, \icarus, 171,
  102

\bibitem[{{Christou}(2000)}]{Christou_2000}
{Christou}, A.~A. 2000, \icarus, 144, 1

\bibitem[{{{\'C}uk} {et~al.}(2012){{\'C}uk}, {Hamilton}, \&
  {Holman}}]{Cuk_etal_2012}
{{\'C}uk}, M., {Hamilton}, D.~P., \& {Holman}, M.~J. 2012, \mnras, 426, 3051

\bibitem[{{de la Fuente Marcos} \& {de la Fuente
  Marcos}(2013)}]{delaFuenteMarcos_delaFuenteMarcos_2013}
{de la Fuente Marcos}, C., \& {de la Fuente Marcos}, R. 2013, \mnras, 432, 886

\bibitem[{{de la Fuente Marcos} \& {de la Fuente
  Marcos}(2014)}]{delaFuenteMarcos_delaFuenteMarcos_2014}
---. 2014, \mnras, 439, 2970

\bibitem[{{de la Fuente Marcos} \& {de la Fuente
  Marcos}(2017)}]{delaFuenteMarcos_delaFuenteMarcos_2017}
---. 2017, Research Notes of the American Astronomical Society, 1, 3

\bibitem[{{Dermott} {et~al.}(1994){Dermott}, {Jayaraman}, {Xu}, {Gustafson}, \&
  {Liou}}]{Dermott_etal_1994}
{Dermott}, S.~F., {Jayaraman}, S., {Xu}, Y.~L., {Gustafson}, B.~{\AA}.~S., \&
  {Liou}, J.~C. 1994, \nat, 369, 719

\bibitem[{{Devog{\`e}le} {et~al.}(2019){Devog{\`e}le}, {Moskovitz}, {Thirouin},
  {Gustaffson}, {Magnuson}, {Thomas}, {Willman}, {Christensen}, {Person},
  {Binzel}, {Polishook}, {DeMeo}, {Hinkle}, {Trilling}, {Mommert}, {Burt}, \&
  {Skiff}}]{Devogele_etal_2019}
{Devog{\`e}le}, M., {Moskovitz}, N., {Thirouin}, A., {et~al.} 2019, \aj, 158,
  196

\bibitem[{{Dohnanyi}(1969)}]{Dohnanyi_1969}
{Dohnanyi}, J.~S. 1969, \jgr, 74, 2531

\bibitem[{{Drlica-Wagner} {et~al.}(2018){Drlica-Wagner}, {Sevilla-Noarbe},
  {Rykoff}, {Gruendl}, {Yanny}, {Tucker}, {Hoyle}, {Carnero Rosell},
  {Bernstein}, {Bechtol}, {Becker}, {Benoit-L{\'e}vy}, {Bertin}, {Carrasco
  Kind}, {Davis}, {de Vicente}, {Diehl}, {Gruen}, {Hartley}, {Leistedt}, {Li},
  {Marshall}, {Neilsen}, {Rau}, {Sheldon}, {Smith}, {Troxel}, {Wyatt}, {Zhang},
  {Abbott}, {Abdalla}, {Allam}, {Banerji}, {Brooks}, {Buckley-Geer}, {Burke},
  {Capozzi}, {Carretero}, {Cunha}, {D'Andrea}, {da Costa}, {DePoy}, {Desai},
  {Dietrich}, {Doel}, {Evrard}, {Fausti Neto}, {Flaugher}, {Fosalba},
  {Frieman}, {Garc{\'\i}a-Bellido}, {Gerdes}, {Giannantonio}, {Gschwend},
  {Gutierrez}, {Honscheid}, {James}, {Jeltema}, {Kuehn}, {Kuhlmann},
  {Kuropatkin}, {Lahav}, {Lima}, {Lin}, {Maia}, {Martini}, {McMahon},
  {Melchior}, {Menanteau}, {Miquel}, {Nichol}, {Ogand o}, {Plazas}, {Romer},
  {Roodman}, {Sanchez}, {Scarpine}, {Schindler}, {Schubnell}, {Smith}, {Smith},
  {Soares-Santos}, {Sobreira}, {Suchyta}, {Tarle}, {Vikram}, {Walker},
  {Wechsler}, {Zuntz}, \& {DES Collaboration}}]{Drlica_etal_2018}
{Drlica-Wagner}, A., {Sevilla-Noarbe}, I., {Rykoff}, E.~S., {et~al.} 2018,
  \apjs, 235, 33

\bibitem[{{Flaugher} {et~al.}(2015){Flaugher}, {Diehl}, {Honscheid}, {Abbott},
  {Alvarez}, {Angstadt}, {Annis}, {Antonik}, {Ballester}, {Beaufore},
  {Bernstein}, {Bernstein}, {Bigelow}, {Bonati}, {Boprie}, {Brooks},
  {Buckley-Geer}, {Campa}, {Cardiel-Sas}, {Castand er}, {Castilla}, {Cease},
  {Cela-Ruiz}, {Chappa}, {Chi}, {Cooper}, {da Costa}, {Dede}, {Derylo},
  {DePoy}, {de Vicente}, {Doel}, {Drlica-Wagner}, {Eiting}, {Elliott}, {Emes},
  {Estrada}, {Fausti Neto}, {Finley}, {Flores}, {Frieman}, {Gerdes},
  {Gladders}, {Gregory}, {Gutierrez}, {Hao}, {Holland}, {Holm}, {Huffman},
  {Jackson}, {James}, {Jonas}, {Karcher}, {Karliner}, {Kent}, {Kessler},
  {Kozlovsky}, {Kron}, {Kubik}, {Kuehn}, {Kuhlmann}, {Kuk}, {Lahav}, {Lathrop},
  {Lee}, {Levi}, {Lewis}, {Li}, {Mand richenko}, {Marshall}, {Martinez},
  {Merritt}, {Miquel}, {Mu{\~n}oz}, {Neilsen}, {Nichol}, {Nord}, {Ogando},
  {Olsen}, {Palaio}, {Patton}, {Peoples}, {Plazas}, {Rauch}, {Reil}, {Rheault},
  {Roe}, {Rogers}, {Roodman}, {Sanchez}, {Scarpine}, {Schindler}, {Schmidt},
  {Schmitt}, {Schubnell}, {Schultz}, {Schurter}, {Scott}, {Serrano}, {Shaw},
  {Smith}, {Soares-Santos}, {Stefanik}, {Stuermer}, {Suchyta}, {Sypniewski},
  {Tarle}, {Thaler}, {Tighe}, {Tran}, {Tucker}, {Walker}, {Wang}, {Watson},
  {Weaverdyck}, {Wester}, {Woods}, {Yanny}, \& {DES
  Collaboration}}]{Flaugher_etal_2015}
{Flaugher}, B., {Diehl}, H.~T., {Honscheid}, K., {et~al.} 2015, \aj, 150, 150

\bibitem[{{Granvik} {et~al.}(2018){Granvik}, {Morbidelli}, {Jedicke}, {Bolin},
  {Bottke}, {Beshore}, {Vokrouhlick{\'y}}, {Nesvorn{\'y}}, \&
  {Michel}}]{Granvik_etal_2018}
{Granvik}, M., {Morbidelli}, A., {Jedicke}, R., {et~al.} 2018, \icarus, 312,
  181

\bibitem[{{Ivezi{\'c}} {et~al.}(2001){Ivezi{\'c}}, {Tabachnik}, {Rafikov},
  {Lupton}, {Quinn}, {Hammergren}, {Eyer}, {Chu}, {Armstrong}, {Fan},
  {Finlator}, {Geballe}, {Gunn}, {Hennessy}, {Knapp}, {Leggett}, {Munn},
  {Pier}, {Rockosi}, {Schneider}, {Strauss}, {Yanny}, {Brinkmann}, {Csabai},
  {Hindsley}, {Kent}, {Lamb}, {Margon}, {McKay}, {Smith}, {Waddel}, {York}, \&
  {SDSS Collaboration}}]{Ivezic_etal_2001}
{Ivezi{\'c}}, {\v{Z}}., {Tabachnik}, S., {Rafikov}, R., {et~al.} 2001, \aj,
  122, 2749

\bibitem[{{Jedicke} {et~al.}(2016){Jedicke}, {Bolin}, {Granvik}, \&
  {Beshore}}]{Jedicke_etal_2016}
{Jedicke}, R., {Bolin}, B., {Granvik}, M., \& {Beshore}, E. 2016, \icarus, 266,
  173

\bibitem[{{Jones} {et~al.}(2013){Jones}, {Bewsher}, \&
  {Brown}}]{Jones_etal_2013}
{Jones}, M.~H., {Bewsher}, D., \& {Brown}, D.~S. 2013, Science, 342, 960

\bibitem[{{Leinert} \& {Moster}(2007)}]{Leinert_Moster_2007}
{Leinert}, C., \& {Moster}, B. 2007, \aap, 472, 335

\bibitem[{{Levison} \& {Duncan}(2013)}]{Levison_Duncan_2013}
{Levison}, H.~F., \& {Duncan}, M.~J. 2013, {SWIFT: A solar system integration
  software package}, Astrophysics Source Code Library, , , ascl:1303.001

\bibitem[{{Masiero} {et~al.}(2009){Masiero}, {Jedicke}, {{\v{D}}urech}, {Gwyn},
  {Denneau}, \& {Larsen}}]{Masiero_etal_2009}
{Masiero}, J., {Jedicke}, R., {{\v{D}}urech}, J., {et~al.} 2009, \icarus, 204,
  145

\bibitem[{{Mikkola} {et~al.}(2004){Mikkola}, {Brasser}, {Wiegert}, \&
  {Innanen}}]{Mikkola_etal_2004}
{Mikkola}, S., {Brasser}, R., {Wiegert}, P., \& {Innanen}, K. 2004, \mnras,
  351, L63

\bibitem[{{Mobasher}(2002)}]{Mobasher_2002}
{Mobasher}, B. 2002, {HST Data Handbook: Introduction to Reducing HST Data,
  Volume 1, Version 4.0}

\bibitem[{{Morais} \& {Morbidelli}(2006)}]{Morais_Morbidelli_2006}
{Morais}, M.~H.~M., \& {Morbidelli}, A. 2006, \icarus, 185, 29

\bibitem[{{Muinonen} {et~al.}(2010){Muinonen}, {Belskaya}, {Cellino},
  {Delb{\`o}}, {Levasseur-Regourd}, {Penttil{\"a}}, \&
  {Tedesco}}]{Muinonen_etal_2010}
{Muinonen}, K., {Belskaya}, I.~N., {Cellino}, A., {et~al.} 2010, \icarus, 209,
  542

\bibitem[{{Oszkiewicz} {et~al.}(2012){Oszkiewicz}, {Bowell}, {Wasserman},
  {Muinonen}, {Penttil{\"a}}, {Pieniluoma}, {Trilling}, \&
  {Thomas}}]{Oszkieweicz_etal_2012}
{Oszkiewicz}, D.~A., {Bowell}, E., {Wasserman}, L.~H., {et~al.} 2012, \icarus,
  219, 283

\bibitem[{{Penttil{\"a}} {et~al.}(2016){Penttil{\"a}}, {Shevchenko}, {Wilkman},
  \& {Muinonen}}]{Penttila_etal_2016}
{Penttil{\"a}}, A., {Shevchenko}, V.~G., {Wilkman}, O., \& {Muinonen}, K. 2016,
  \planss, 123, 117

\bibitem[{{Pokorn{\'y}} \& {Kuchner}(2019)}]{Pokorny_Kuchner_2019}
{Pokorn{\'y}}, P., \& {Kuchner}, M. 2019, \apjl, 873, L16

\bibitem[{{Sheppard} \& {Trujillo}(2009)}]{SHEPPARD_TRUJILLO_2009}
{Sheppard}, S.~S., \& {Trujillo}, C.~A. 2009, \icarus, 202, 12

\bibitem[{{Sheppard} {et~al.}(2019){Sheppard}, {Trujillo}, {Tholen}, \&
  {Kaib}}]{Sheppard_etal_2019}
{Sheppard}, S.~S., {Trujillo}, C.~A., {Tholen}, D.~J., \& {Kaib}, N. 2019, \aj,
  157, 139

\bibitem[{{Shevchenko} {et~al.}(2016){Shevchenko}, {Belskaya}, {Muinonen},
  {Penttil{\"a}}, {Krugly}, {Velichko}, {Chiorny}, {Slyusarev}, {Gaftonyuk}, \&
  {Tereschenko}}]{Shevchenko_etal_2016}
{Shevchenko}, V.~G., {Belskaya}, I.~N., {Muinonen}, K., {et~al.} 2016, \planss,
  123, 101

\bibitem[{{Shevchenko} {et~al.}(2019){Shevchenko}, {Belskaya}, {Mikhalchenko},
  {Muinonen}, {Penttil{\"a}}, {Gritsevich}, {Shkuratov}, {Slyusarev}, \&
  {Videen}}]{Shevchenko_etal_2019}
{Shevchenko}, V.~G., {Belskaya}, I.~N., {Mikhalchenko}, O.~I., {et~al.} 2019,
  \aap, 626, A87

\bibitem[{{Smith} {et~al.}(2002){Smith}, {Tucker}, {Kent}, {Richmond},
  {Fukugita}, {Ichikawa}, {Ichikawa}, {Jorgensen}, {Uomoto}, {Gunn}, {Hamabe},
  {Watanabe}, {Tolea}, {Henden}, {Annis}, {Pier}, {McKay}, {Brinkmann}, {Chen},
  {Holtzman}, {Shimasaku}, \& {York}}]{Smith_etal_2002}
{Smith}, J.~A., {Tucker}, D.~L., {Kent}, S., {et~al.} 2002, \aj, 123, 2121

\bibitem[{{Trujillo} \& {Jewitt}(1998)}]{Trujillo_Jewitt_1998}
{Trujillo}, C., \& {Jewitt}, D. 1998, \aj, 115, 1680

\bibitem[{{Waszczak} {et~al.}(2015){Waszczak}, {Chang}, {Ofek}, {Laher},
  {Masci}, {Levitan}, {Surace}, {Cheng}, {Ip}, {Kinoshita}, {Helou}, {Prince},
  \& {Kulkarni}}]{Waszcak_etal_2015}
{Waszczak}, A., {Chang}, C.-K., {Ofek}, E.~O., {et~al.} 2015, \aj, 150, 75

\bibitem[{{Wiegert} {et~al.}(2000){Wiegert}, {Innanen}, \&
  {Mikkola}}]{Wiegert_etal_2000}
{Wiegert}, P., {Innanen}, K., \& {Mikkola}, S. 2000, \icarus, 145, 33

\bibitem[{{Ye} {et~al.}(2020){Ye}, {Masci}, {Ip}, {Prince}, {Helou},
  {Farnocchia}, {Bellm}, {Dekany}, {Graham}, {Kulkarni}, {Kupfer}, {Mahabal},
  {Ngeow}, {Reiley}, \& {Soumagnac}}]{Ye_etal_2020}
{Ye}, Q., {Masci}, F.~J., {Ip}, W.-H., {et~al.} 2020, \aj, 159, 70

\bibitem[{{Zappala} {et~al.}(1990){Zappala}, {Cellino}, {Barucci},
  {Fulchignoni}, \& {Lupishko}}]{Zappala_etal_1990}
{Zappala}, V., {Cellino}, A., {Barucci}, A.~M., {Fulchignoni}, M., \&
  {Lupishko}, D.~F. 1990, \aap, 231, 548

\bibitem[{{Zavodny} {et~al.}(2008){Zavodny}, {Jedicke}, {Beshore}, {Bernardi},
  \& {Larson}}]{Zavodny_etal_2008}
{Zavodny}, M., {Jedicke}, R., {Beshore}, E.~C., {Bernardi}, F., \& {Larson}, S.
  2008, \icarus, 198, 284

\end{thebibliography}

\end{document}